\documentclass[prd,nofootinbib]{revtex4}
\usepackage{graphicx, epsfig}
\usepackage{color}
\usepackage{mathrsfs}
\usepackage{amsmath}

\newcommand{\be}{\begin{equation}}
\newcommand{\ee}{\end{equation}}
\newcommand{\bd}{\begin{equation*}}
\newcommand{\ed}{\end{equation*}}
\newcommand{\bea}{\begin{eqnarray}}
\newcommand{\eea}{\end{eqnarray}}

\newcommand{\gapp}{\mathrel{\raise.3ex\hbox{$>$}\mkern-14mu
              \lower0.6ex\hbox{$\sim$}}}
\newcommand{\lapp}{\mathrel{\raise.3ex\hbox{$<$}\mkern-14mu
              \lower0.6ex\hbox{$\sim$}}}

\begin{document}

\title{Quantum Collapse of a Charged $n$-dimensional BTZ-like Domain Wall}
\author{Eric Greenwood}
\affiliation{Department of Geology and Physics, University of Southern Indiana, Evansville, IN 47712}
\begin{abstract}

We investigate both the classical and quantum gravitational collapse of a charged, non-rotating 
$n$-dimensional BTZ black hole in AdS space. This is done by first deriving the conserved mass of 
a ``spherically" symmetric domain wall, which is then taken as the classical Hamiltonian of the domain 
wall. 

In the classical picture, we show that, as far as the asymptotic observer is concerned, the details 
of the collapse depend on the amount of charge present in the domain wall. In the both the 
extremal and non-extremal cases, the collapse takes an infinite amount of observer time to complete. 
However, in the over-charged case, the collapse never actually occurs, instead one finds an 
oscillatory solution which prevents the formation of a naked singularity. As far as the infalling 
observer is concerned, the collapse is completed within a finite amount of proper time. Thus, the 
gravitational collapse follows that of the typical formation of a black hole via gravitational collapse. 

Quantum mechanically, in the absence of induced quasi-particle production and fluctuations of the 
metric geometry, we take primary interest in the behavior of the collapse near the horizon and 
near the classical singularity from the point of view of both asymptotic and infalling observers. For 
the asymptotic observer, we find that quantum effects near the horizon do not change the classical 
conclusions. For the infalling observer, the most interesting quantum mechanical effect are when the 
collapsing shell approaches the classical singularity. Here, we find, that the quantum effects in this 
region display non-local effects, which depend on the energy density of the domain wall. 

\end{abstract}

\maketitle

\section{Introduction}

One of the most studied scenarios in theoretical physics is that of gravitational collapse via the 
collapse of a shell of matter, since the scenario has a vast number of theoretical applications. For 
example, the process of gravitational collapse has been used in many different subfields of theoretical 
physics, whether it be to study the classical formation of a massive black hole 
\cite{Oppenheimer:1939ue,Vachaspati:2006ki,Greenwood:2008ht,Greenwood:2009gp}, quantum 
formation of a massive black hole \cite{Saini:2014qpa}, induced quasi-particle production during the 
collapse \cite{Greenwood:2008zg}, formation of Hawking radiation \cite{Hawking:1976ra}, or 
thermalization processes \cite{Danielsson:1999zt,Danielsson:1999fa,Baron:2012fv} within the context 
of the 
AdS/CFT correspondence and the Functional Schr\"odinger formalism. However, not only is the 
gravitational collapse a matter shell important, but the presence of additional quantum numbers, such 
as charge and angular momentum, can play a great role in the collapse as well
\cite{Lopez:1988gt,Wang:2009ay,Cohen:1968}, as well as lead to additional interesting effects during 
the collapse. 

Moreover, due to the applications within the AdS/CFT correspondence, gravitational collapse in AdS 
has become of greater importance. Recently, the gravitational collapse of an $n$-dimensional BTZ 
matter shell was studied in \cite{Greenwood:2015xwa}, however the study of the collapse in the 
presence of additional quantum numbers is still required. For example, according to the AdS/CFT 
dictionary, the presence of charge on the collapse shell; that is, within the bulk geometry, corresponds 
to the presence of a chemical potential and charge density on the boundary and hence in the field 
theory \cite{Maldacena:1997zz,Hartnoll:2008vx}. Therefore, it is worth investigating the gravitational 
collapse of an $n$-dimensional, charged, BTZ matter shell to gain further insight into the collapse 
process. In this paper, we thus investigate both the classical and quantum collapse of the above 
scenario. 

\section{Setup}
\label{sec:Setup}

To determine the equations of motion for the $n$-dimensional collapse, we will consider the collapse of 
an infinitely thin domain wall, which represents a shell of charged matter. Moreover, we consider 
``spherical" domain walls, so that the wall is completely described by the radial degree of freedom, 
$R(t)$. Using the Gauss-Codazzi equations, the conserved mass may be determined by considering 
the an $(n-1)$-dimensional metric that is intrinsic to a hypersurface $S$, which contains a jump 
discontinuity, see \cite{Ipser:1983db} for details. However, it was show in \cite{Greenwood:2015xwa} 
that for a general `spherically" symmetric collapse, the conserved mass may be easily determined by 
solving the expression 
\be
  \alpha-\beta=\frac{8\pi\sigma R}{n-2},
  \label{EOM}
\ee
where $\sigma$ is the energy density of the domain wall, $\alpha$ and $\beta$ are defined from the 
metric coefficients:
\begin{align}
  \alpha&\equiv ft_{-,\tau}=\sqrt{f+R_{\tau}^2}\label{alpha_BTZ}\\
  \beta&\equiv Ft_{+,\tau}=\sqrt{F+R_\tau^2}\label{beta_BTZ}
\end{align}
and the exterior $(+)$ and interior $(-)$ metrics are given by,
\begin{align*}
  (ds^2)_+&=-F(r)dt_+^2+\frac{1}{F(r)}dr^2+r^2d\Omega_{n-2}\\
  (ds^2)_-&=-f(r)dt_-^2+\frac{1}{f(r)}dr^2+r^2d\Omega_{n-2}
\end{align*}
In (\ref{alpha_BTZ}) and (\ref{beta_BTZ}), $\tau$ is the proper time of an observer moving with 
$n$-velocity $u^a$ at the wall, hence $t_{\pm,\tau}$ is the propertime derivative of the coordinate 
time of the exterior and interior metrics. 

What is interesting about this result is that (\ref{EOM}) only depends on the surface density of the 
collapsing shell, not on the tension. That is, the solution for the collapsing shell is the same for any 
equation of state, up to the spatial dependence of the surface density. That is, even though we are 
considering the collapse of a domain wall, the solution is generic for different types of matter. 

In the present paper, we are interested in the collapse of a massive, charged, $n$-dimensional BTZ 
domain wall. From \cite{Hendi:2010px,Hendi:2012zz}, we take that the exterior metric coefficient $F$ 
is given as
\be
  F=\frac{R^2}{\ell^2}-\frac{M}{R^{n-3}}-\frac{2^{(n-1)/2}}{R^{n-3}}Q^{n-1}\ln\frac{R}{\ell}.
  \label{F}
\ee
and the interior metric coefficient $f$ is given as
\be
  f=\frac{R^2}{\ell^2}.
  \label{f}
\ee
From the metric coefficients, using (\ref{EOM}), we can solve for the mass, which is then given by
\begin{align}
  M&=\frac{16\pi\sigma R^{n-2}}{n-2}\left(\alpha-\frac{4\pi\sigma R}{n-2}\right)-2^{(n-1)/2}Q^{n-1}\ln\frac{R}{\ell}\nonumber\\
     &=\frac{16\pi\sigma R^{n-2}\sqrt{f}}{n-2}\left(\sqrt{1+\frac{R_\tau^2}{f}}-\frac{4\pi\sigma R}{(n-2)\sqrt{f}}\right)-2^{(n-1)/2}Q^{n-1}\ln\frac{R}{\ell}
  \label{BTZ_mass}
\end{align}
where in the last line we have rewritten the mass for later convenience. As mentioned previously, the 
Gauss-Codazzi equations lead to a conserved mass a conserved mass. In Appendix \ref{ch:check}, we 
show that (\ref{BTZ_mass}) is indeed a constant of motion.  

Since the mass in (\ref{BTZ_mass}) is conserved, the mass then represents the total energy of the 
collapsing shell and may be treated as the Hamiltonian of the system. This conclusion makes 
sense from the structure of the conserved mass. To see this, notice that for a static domain wall, 
$R_\tau=0$, the first term in the brackets is just the total rest mass, the second term is the binding 
energy, and the last term is the electromagnetic contribution. Hence, for the non-static domain wall, 
$R_\tau\not=0$, the first term in the brackets now accounts for the relativistic kinetic energy. Therefore, 
we will determine the equations of motion for the collapsing shell using the conserved mass in 
(\ref{BTZ_mass}) as the Hamiltonian system. 

As is well known, since the Hamiltonian is not invariant under coordinate system transformations, we 
will work with the Lagrangian for the system. The Lagrangian is given by,
\begin{align}
  L(\tau)&=-\frac{16\pi\sigma R^{n-2}}{n-2}\left(\alpha-\frac{4\pi\sigma R}{n-2}-R_\tau\sinh^{-1}\sqrt{\frac{R_\tau^2}{f}}\right)+2^{(n-1)/2}Q^{n-1}\ln\frac{R}{\ell}\nonumber\\
     &=-\frac{16\pi\sigma R^{n-2}}{n-2}\left(\sqrt{f+R_\tau^2}-\frac{4\pi\sigma R}{n-2}-R_\tau\sinh^{-1}\sqrt{\frac{R_\tau^2}{f}}\right)+2^{(n-1)/2}Q^{n-1}\ln\frac{R}{\ell}.
  \label{Lm}
\end{align}
For later convenience, we note that the generalized momentum associated with (\ref{Lm}) is given by
\be
  P(\tau)=\frac{16\pi\sigma R^{n-2}}{n-2}\sinh^{-1}\sqrt{\frac{R_\tau^2}{f}}.
  \label{P_tau}
\ee

Let's now the classical and quantum equations of motion for the asymptotic observer. 

\section{Asymptotic Observer}
\label{sec:Asymptotic}

Before we can obtain the equations of motion for the asymptotic observer, we must first transform 
the Lagrangian in (\ref{Lm}) to the coordinate time $t_+$, which may be done by considering the 
effective action
\bd
  S(\tau)=\int d\tau L_m(\tau).
\ed
Using (\ref{beta_BTZ}), in the $t_+$ coordinate, the Lagrangian (\ref{Lm}) takes the form
\be
  L(t)=-\frac{16\pi\sigma R^{n-2}}{n-2}\left(\sqrt{\frac{fF^2}{\beta^2}+\dot R^2}-\frac{4\pi\sigma RF}{\beta(n-2)}-\dot R\sinh^{-1}\left(\frac{\beta}{F}\sqrt{\frac{\dot R^2}{f}}\right)\right)+\frac{2^{(n-1)/2}Q^{n-1}F}{\beta}\ln\frac{R}{\ell}
  \label{Lm_t}
\ee
where $\dot R=dR/dt$ and $\beta$ is given in (\ref{beta_BTZ}). 
The Hamiltonian and generalized momentum in the asymptotic observer time coordinate may then 
be obtained from (\ref{Lm_t}) in the usual manner. 

For the purposes of the present paper, we are, however, only interested in the near horizon regime of 
the collapse; that is, the last moments of the collapse before the formation of the black hole. Thus, we 
are only interested in the $R\sim R_+$ regime of the collapse, where $R_+$ is the horizon radius. 
In terms of the metric coefficient, this corresponds to when the exterior metric coefficient $F\sim0$. 
Notice that in this regime, the interior metric coefficient $f$ is simply a constant, since $f=R_+^2/\ell^2$. 
In this limit, the generalized momentum takes the form
\be
  P(t)\approx\frac{16\pi\mu R^{n-2}\dot R}{(n-2)\sqrt{F}\sqrt{F^2-\dot R^2}},
  \label{P_t approx}
\ee
where 
\bd
  \mu\equiv\sigma\left(\sqrt{f}-\frac{4\pi\sigma R_H}{n-2}-\frac{(n-2)2^{(n+1)/2}Q^{n+1}}{16\pi\sigma R^{n-2}_H}\ln\frac{R_H}{\ell}\right)
\ed
is the modified energy density of the domain wall and $f$ is evaluated at $R=R_+$. In this limit, we may 
also obtain the Hamiltonian 
\be
  H(t)\approx\frac{16\pi\mu R^{n-2}F^{3/2}}{(n-2)\sqrt{F^2-\dot R^2}}.
  \label{H_t approx}
\ee
To determine the quantum equation of motion, it is convenient to invert (\ref{P_t approx}) so that we 
may rewrite (\ref{H_t approx}) in terms of the generalized momentum:
\be
  H=\sqrt{(FP)^2+\left(\frac{16\pi\mu R}{n-2}\right)^2}.
  \label{H_t P}
\ee
We can see that (\ref{H_t P}) has the form of the Hamiltonian of a relativistic particle with a position 
dependent mass term. 

We are now in a position to determine the classical and quantum equations of motion for the collapse. 
Let us determine the classical equations of motion for the collapse first. 

\subsection{Classical Equations of Motion}

As the Hamiltonian is a conserved quantity, i.e.~a constant, we can determine the velocity of 
the domain wall. From (\ref{H_t approx}) we have
\be
  \dot R=\pm F\sqrt{1-\frac{FR^{2(n-2)}}{h^2}},
\ee
where $h\equiv H(n-2)/16\pi\mu$, or in the near horizon limit becomes
\bd
  \dot R\approx\pm F\left(1-\frac{1}{2}\frac{FR^{2(n-2)}}{h^2}\right)\approx\pm F.
\ed
Hence, we can see that the dynamics of the collapse, in the late-time limit, are solely governed by 
the exterior metric coefficient, $\dot R\approx-F$, where the negative sign is chosen due to the fact 
that we are interested in collapse of the domain wall. 

Since the domain wall is charged, there are three interesting cases to look at: The non-extremal case, 
the extremal case and the overcharged (or naked singularity) case. Let's consider the dynamics of 
each of these three cases.

\subsubsection{Non-Extremal Case}

In the non-extremal case, the black hole has two horizons, an inner and outer horizon 
\cite{Hendi:2010px}. For arbitrary dimension $n$, the horizons are located at 
\begin{align*}
  R_-&=\ell\exp\left[-\frac{1}{n-1}W\left(-\frac{(n-1)\ell^{n-3}}{2^{(n-1)/2}Q^{n-1}}e^{-\frac{(n-1)M}{2^{(n-1)/2}Q^{n-1}}}\right)-\frac{M}{2^{(n-1)/2}Q^{n-1}}\right],\\
  R_+&=\ell\exp\left[-\frac{1}{n-1}W_{-1}\left(-\frac{(n-1)\ell^{n-3}}{2^{(n-1)/2}Q^{n-1}}e^{-\frac{(n-1)M}{2^{(n-1)/2}Q^{n-1}}}\right)-\frac{M}{2^{(n-1)/2}Q^{n-1}}\right],
\end{align*}
where $W$ is the {\it LambertW} function\footnote{The {\it LambertW} function is also known as the 
omega function or product logarithm.}, which satisfies $W(x)e^{W(x)}=x$, which are solutions 
to 
\bd
  F(R_+)=\frac{R_+^2}{\ell^2}-\frac{1}{R_+^{n-3}}\left(M+2^{(n-1)/2}Q^{n-1}\ln\frac{R_+}{\ell}\right)=0
\ed
and $R_+$ is the largest real root. For real values of the argument, $W$ has only two real branches 
which are single valued: $W_0(x)$ being the principle branch is single-valued for $W\geq-1$ and 
$W_{-1}(x)$ being the lower branch which is single-valued for $W\leq-1$. 

To solve the equation of motion, we first note that the metric coefficient may be written as 
\begin{align}
  F&=\frac{\sqrt2QR^n-\sqrt2\ell^2MQR+2^{n/2}\ell^2Q^n\ln(R/\ell)}{\sqrt2\ell^2QR^{n-2}}\nonumber\\
     &\equiv\frac{(R-R_+)(R-R_-)C}{\sqrt2\ell^2QR^{n-2}}\label{F_decomp},
\end{align}
where $C$ are all other roots. Hence, as the domain wall approaches the outer horizon, the metric 
coefficient goes to 
\bd
  F\to(R-R_+)\frac{(R_+-R_-)C_+}{\sqrt2\ell^2QR_+^{n-2}}\equiv(R-R_+)A,
\ed
where $A$ is just a constant and $C_+$ denotes that all other roots are evaluated at $R=R_+$. Thus, 
as the domain wall approaches the outer horizon, the position of the domain wall is given by
\be
  R\approx R_++(R_0-R_+)e^{\pm At_+}=R_++(R_0-R_+)e^{\pm\frac{(R_+-R_-)C_+t_+}{\sqrt2\ell^2QR_+^{n-2}}}
  \label{R_asymp}
\ee
where $R_0$ is the initial position of the domain wall; that is, as far as the asymptotic observer is 
concerned, it takes an infinite amount of observer time for the domain wall to complete the collapse. 

\subsubsection{Extremal Case}

In the extremal case, the inner and outer radii are the same, hence there is only one real root 
\cite{Hendi:2010px}: $R_-=R_+$. In this case, the metric coefficient (\ref{F}) takes the form
\bd
  F=\frac{(R-R_+)^2}{\sqrt2\ell^2QR^{n-2}}.
\ed
That is, in the vicinity of the horizon, the metric coefficient may be written as 
\bd
  F=\frac{(R-R_+)^2}{\sqrt2\ell^2QR_+^{n-2}}\equiv(R-R_+)^2B_+,
\ed
where $B_+$ is just a constant. Substituting this into the equation of motion for the domain wall, we 
then find that the position of the domain wall is now given by
\bd
  R\approx R_++\frac{R_0-R_+}{1\pm B_+(R_0-R_+)t_+}.
\ed
Again we see that the domain wall only reaches the horizon after an infinite amount of observer 
time. 

\subsubsection{Over-charged Case}

In the case where domain wall is over-charged, hence the charge is more important than the mass, 
as a result, both the inner and outer radii become imaginary. 
Thus, if the collapse were to proceed all the way to the outer, imaginary, horizon, this would represent 
a violation of the cosmic censorship conjecture and result in a naked singularity. However, we 
can see that this does not happen here: As we can see from (\ref{R_asymp}), the exponent in the 
exponential will be complex and hence we will have an oscillatory solution. This implies that the domain 
wall will collapse to a minimal radius due to the gravitational attraction, however the electromagnetic 
repulsion will then overcome the gravitational attraction and generate a bouncing solution. As the 
domain wall expands, the gravitational attraction will once again dominate and cause the domain 
wall to once again collapse, to which the process will be repeated. 

\subsection{Quantum Equations of Motion}
\label{sec:QuantumAsymptotic}

For the quantum solution, we will only be concerned with the non-extremal case.

From (\ref{H_t P}), we see that the quantum Hamiltonian, as far as the asymptotic observer is 
concerned, is the same as the quantum Hamiltonian found in 
\cite{Vachaspati:2006ki,Greenwood:2008zg,Greenwood:2009gp}. As a result, we can easily write the 
solution as a Gaussian wave-packet solution which is squeezing\footnote{Here, the Gaussian wave 
packet is squeezing as a result of the relationship between $R$ and $u$, which is given by $dR=fdu$.} 
during its approach to the horizon
\bd
  \Psi=\frac{1}{\sqrt{2\pi}s}e^{-(u+t)^2/2s^2},
\ed
where 
\bd
  u=\int\frac{dR}{F}
\ed
is the tortoise coordinate and $s$ is the width of the wave packet. In the $u$-coordinate, one can 
easily see that the horizon is moved to to infinity; that is $u(R_+)=-\infty$. That is, since the wave 
packet must travel and infinite distance to reach the horizon, it also takes an infinite amount of time for 
the wave packet to reach the horizon. Therefore, this result does not contradict the classical result of 
taking an infinite amount of observer time to collapse to form a black hole.

\section{Infalling Observer}
\label{sec:Infalling}

Let's now turn our attention to the infalling observer, whose Hamiltonian is given in (\ref{BTZ_mass}) 
and generalized momentum is given in (\ref{P_tau}). 

\subsection{Classical Equations of Motion}

Just like the asymptotic observer, we are interested in the classical equation of motion for the 
collapsing domain wall in the vicinity of the horizon. Classically, one should expect that the horizon 
is not an obstacle for the infalling observer since the horizon only represents a coordinate singularity 
and not a space-time singularity. 

We may obtain the velocity as far as the infalling observer is concerned, of the domain wall from 
(\ref{BTZ_mass}),
\be
  R_\tau=\sqrt{\left(\frac{\tilde h+2^{(n-1)/2}Q^n\ln(R/\ell)}{R^{n-2}}+\frac{4\pi\sigma R}{n-2}\right)^2-f},
  \label{R_tau}
\ee
where $\tilde h=(n-2)H(\tau)/8\pi\sigma$. Due to the presence of the natural log, (\ref{R_tau}) is not 
easily solved in closed form. Thus, we will seek an approximate solution to (\ref{R_tau}). As 
a zeroth order approximation, we can see that (\ref{R_tau}) is just a constant in the near horizon region, 
$R\sim R_+$, so we may easily obtain the zeroth order solution, which is given by 
\bd
  R(\tau)=R_0-\tau\sqrt{\left(\frac{\tilde h+2^{(n-1)/2}Q^n\ln(R_+/\ell)}{R_+^{n-2}}+\frac{4\pi\sigma R_+}{n-2}\right)^2-\frac{R_+^2}{\ell^2}}
\ed
where again the minus sign is chosen due to the collapse of the shell and $R_0$ is again the initial 
position of the collapsing domain wall at $\tau=0$. As is expected, the infalling observer will see the 
horizon formed in a finite amount of time, which, as stated previously, is expected.

%
\subsection{Quantum Equations of Motion}
\label{sec:QuantumInfalling}

As far as the infalling observer is concerned, the more interesting limit is the near singularity limit of the 
collapse, hence the $R\to0$ region. To obtain the quantum Hamiltonian of the system, we start with 
(\ref{P_tau}), which may be inverted so that we may rewrite the Hamiltonian (\ref{BTZ_mass}) in 
terms of the generalized momentum. As a result, (\ref{BTZ_mass}) takes the form.
\be
  H(\tau)=\frac{16\pi\sigma R^{n-2}}{n-2}\left(\sqrt{f}\cosh\frac{P(\tau)}{8\pi\sigma R}-\frac{4\pi\sigma R}{n-2}\right)-2^{(n-1)/2}Q^{n-1}\ln\frac{R}{\ell}.
  \label{H_tau P}
\ee
Since the velocity is negative near the classical singularity the Hamiltonian, 
written in terms of the conjugate momentum, may be written as
\begin{align}
  H&=\frac{16\pi\sigma R^{n-2}}{n-2}\sqrt{f}\cosh\frac{P(\tau)}{8\pi\sigma R}-2^{(n-1)/2}Q^{n-1}\ln\frac{R}{\ell}\nonumber\\
     &=\frac{8\pi\sigma R^{n-2}}{n-2}\sqrt{f}e^{-\frac{P(\tau)}{8\pi\sigma R}}-2^{(n-1)/2}Q^{n-1}\ln\frac{R}{\ell}\nonumber\\
     &=\frac{8\pi\sigma R^{n-2}}{n-2}\sqrt{f}e^{\frac{i}{8\pi\sigma R}\frac{\partial}{\partial R}}-2^{(n-1)/2}Q^{n-1}\ln\frac{R}{\ell}.
  \label{H near}
\end{align}
Investigating the structure of (\ref{H near}), we can see that the first term is just a translation operator, 
which generates an imaginary translation. One way to see this is to define the new variable $z=R^2$, then we can rewrite 
(\ref{H near}) as
\be
  H=\frac{8\pi\sigma}{n-2}z^{n/2-1}\sqrt{f}e^{\frac{i}{4\pi\sigma}\frac{\partial}{\partial z}}-2^{(n-1)/2}Q^{n-1}\ln\frac{\sqrt{z}}{\ell}.
  \label{Hz Near}
\ee
Hence (\ref{Hz Near}) translates wave function by a non-infinitesimal amount: $\frac{i}{4\pi\sigma}$ 
in $z$ and $\sqrt{\frac{i}{4\pi\sigma}}$ in $R$. That is, as the collapsing shell approaches the 
classical singularity, the wave function is related to its value at some distance point: 
$\Psi(R\to\sqrt{\frac{i}{8\pi\sigma}},\tau)$. 

Moreover, we can see that (\ref{H near}) (and (\ref{Hz Near})) is non-local since it depends on an 
infinite number of derivatives, due to the $R^{-1}$ term in the exponential. The presence of this term 
means that (\ref{H near}) may not be truncated after a few derivatives, unlike a local Hamiltonian. 

As we are interested in the behavior of the wave function near the classical singularity, we must make 
further simplifying approximations. Here, we will work in the limit of large energy density, 
$\sigma\to\infty$, so that we may truncate the expansion of the exponential. As a consequence of this 
approximation, we lessen the non-local behavior of the wave function, however we gain insight into 
the overall behavior of the wave function. In the limit $\sigma\to\infty$, the Schr\"odinger equation 
becomes
\be
  \frac{R^{n-2}}{\ell(n-2)}\partial_R\psi+i2^{(n-1)/2}Q^{n-1}\ln\frac{R}{\ell}\psi=\partial_\tau\psi,
  \label{Schrod}
\ee
where we used (\ref{f}). (\ref{Schrod}) has solution
\bd
  \psi=\psi_0e^{A\left(t-\frac{\ell(n-2)}{(n-3)R^{n-3}}\right)}\exp\left[\frac{i(n-2)2^{(n-1)/2}Q^{n-1}(1+(n-3)\ln(R/\ell))}{(n-3)^2}\right],
\ed
where $A$ is an arbitrary constant. Thus, as the wave function approaches the classical singularity, 
the wave function simplifies to 
\be
  \psi=\psi_0\left(\frac{R}{\ell}\right)^{i(n-2)2^{(n-1)/2}Q^{n-1}/(n-3)}e^{A\left(t-\frac{\ell(n-2)}{(n-3)R^{n-3}}\right)}. 
  \label{wave func}
\ee
In this approximation, the phase factor corresponds to an infinite number of oscillations, while the 
exponential term corresponds to a wave that is propagating to the classical singularity. (\ref{wave func}) 
does imply, however, that the wave function vanishes at the classical singularity, which implies that 
the singularity is a place of infinite density.


\section{Conclusion}
\label{sec:conclusion}

In this paper we studied the gravitational collapse of an $n$-dimensional, spherically symmetric, 
charged BTZ black hole, which is represented by an infinitely thin domain wall. Specifically, we 
studied both the classical and quantum collapse of the domain wall by first determining the 
conserved mass of the domain wall. Since the mass is a conserved quantity, we were allowed 
to interpret the conserved mass as the Hamiltonian of the system. Since we have the freedom to 
choose the viewpoint of any observer, we chose to consider the collapse from the viewpoint of the 
most relevant observers, those being the asymptotic and infalling observers. Furthermore, instead of 
considering the whole history of the collapse, we only considered the collapse in specific regions of 
interest, the near horizon (from both viewpoints) and near classical singularity (only for the infalling 
observer's viewpoint) regimes. 

In Section \ref{sec:Asymptotic} we studied the collapse from the viewpoint of an asymptotic observer, 
both classically and quantum mechanically. Classically, we found that, irrespective of the number 
of dimensions, the horizon is only formed after an infinite amount of observer time, as is expected 
from the infinite gravitational redshift associated with the formation of the horizon. In the quantum 
mechanical picture, we found that\footnote{Note that these calculations were preformed the absence 
of Hawking radiation and back reaction on the metric. Including either or both of these aspects 
may alter the conclusions found here.}, just like in the classical case, the horizon is 
only formed after an infinite amount of observer time. Not unexpectedly, this result implies that simply 
quantizing the matter shell (domain wall) does not lead to quantum fluctuations of the horizon that 
would allow the horizon to be formed in a finite amount of observer time.

In Section \ref{sec:Infalling} we studied the collapse from the viewpoint of an infalling observer, 
which is an observer who is riding on the surface of the domain wall and is parameterized by the 
proper time of the domain wall. Classically, we found that in the zeroth order approximation, the horizon 
is formed in a finite amount of proper time. As in the asymptotic observer case, this result is 
independent of the dimensionality. We also investigated the collapse of the domain wall in the 
vicinity of the classical singularity, $R\to0$. Here, we found that the Hamiltonian takes on a form that 
resembles that of a translation operator, (\ref{H near}), and thus the Hamiltonian translates the wave 
function by a non-infinitesimal amount. That is, the wave function at a given point depends on the 
value of the wave function at a distant point. This behavior is a sign of non-locality. More importantly, 
the distant point depends inversely on the energy density of the domain wall, hence the larger the 
energy density, the closer the point is. As the energy density approaches infinity, points are only 
separated by an infinitesimal amount and locality is restored. An alternative way to see the non-local 
behavior of (\ref{H near}) is to again note the structure of the translation operator. As the differential 
operator enters into the Hamiltonian via an exponential, there is a $R^{-1}$ dependence, 
which means that in a Taylor expansion of the exponential near the classical singularity, the higher 
order terms of the expansion become more important, meaning that would would have to keep all 
infinite number of terms to fully define the Taylor expansion. Furthermore, we found that the wave 
function vanishes near the classical singularity, which can be attributed to either an infinite density 
or an infinite electromagnetic repulsion once the domain wall has collapsed to zero radius. 



\appendix

\section{Check that Mass per unit Length is a constant of Motion}
\label{ch:check}

Here we wish to check that (\ref{BTZ_mass}) is a constant of motion. First, note that we 
can rewrite (\ref{BTZ_mass}) as
\bd
  M=\frac{16\pi\sigma R^{n-2}}{n-2}\left(\alpha-\frac{4\pi\sigma R}{n-2}\right)-2^{(n-1)/2}Q^{n-1}\ln\frac{R}{\ell},
\ed
or using (\ref{EOM}) this becomes
\be
  M=R^{n-3}\left(\alpha^2-\beta^2\right)-2^{(n-1)/2}Q^{n-1}\ln\frac{R}{\ell}.
  \label{mass_alpha}
\ee
Taking the proper time derivative of (\ref{mass_alpha}) we obtain
\bd
  M_\tau=(n-3)R^{n-4}R_\tau\left(\alpha^2-\beta^2\right)+2R^{n-3}\left(\alpha\alpha_{\tau}-\beta\beta_{\tau}\right)-2^{(n-1)/2}Q^{n-1}\frac{2R_\tau}{\ell R}.
\ed
Using (\ref{alpha_BTZ}) we have
\bd
  M_\tau=R_\tau\left[(n-3)R^{n-4}\left(f-F\right)+R^{n-3}\left(f'-F'\right)-2^{(n-1)/2}Q^{n-1}\frac{2}{\ell R}\right],
\ed
where the prime refers to derivative with respect to the position of the domain wall, and using (\ref{F}) 
and (\ref{f}) we finally obtain
\bd
  M_\tau=0,
\ed
which proves that the mass is a conserved quantity.


\begin{thebibliography}{99}

\bibitem{Oppenheimer:1939ue} 
  J.~R.~Oppenheimer and H.~Snyder,
  Phys.\ Rev.\  {\bf 56}, 455 (1939).

\bibitem{Vachaspati:2006ki} 
  T.~Vachaspati, D.~Stojkovic and L.~M.~Krauss,
  Phys.\ Rev.\ D {\bf 76}, 024005 (2007)
  [gr-qc/0609024].

\bibitem{Greenwood:2008ht} 
  E.~Greenwood and D.~Stojkovic,
  JHEP {\bf 0806}, 042 (2008)
  [arXiv:0802.4087 [gr-qc]].

\bibitem{Greenwood:2009gp} 
  E.~Greenwood, E.~Halstead and P.~Hao,
  JHEP {\bf 1002}, 044 (2010)
  [arXiv:0912.1860 [gr-qc]].

\bibitem{Saini:2014qpa} 
  A.~Saini and D.~Stojkovic,
  Phys.\ Rev.\ D {\bf 89}, no. 4, 044003 (2014)
  [arXiv:1401.6182 [gr-qc]].

\bibitem{Greenwood:2008zg} 
  E.~Greenwood and D.~Stojkovic,
  JHEP {\bf 0909}, 058 (2009)
  [arXiv:0806.0628 [gr-qc]].

\bibitem{Hawking:1976ra} 
  S.~W.~Hawking,
  Phys.\ Rev.\ D {\bf 14}, 2460 (1976).
  doi:10.1103/PhysRevD.14.2460

\bibitem{Danielsson:1999zt} 
  U.~H.~Danielsson, E.~Keski-Vakkuri and M.~Kruczenski,
  Nucl.\ Phys.\ B {\bf 563}, 279 (1999)
  [hep-th/9905227].

\bibitem{Danielsson:1999fa} 
  U.~H.~Danielsson, E.~Keski-Vakkuri and M.~Kruczenski,
  JHEP {\bf 0002}, 039 (2000)
  [hep-th/9912209].

\bibitem{Baron:2012fv} 
  W.~Baron, D.~Galante and M.~Schvellinger,
  JHEP {\bf 1303}, 070 (2013)
  [arXiv:1212.5234 [hep-th]].

\bibitem{Wang:2009ay} 
  J.~E.~Wang, E.~Greenwood and D.~Stojkovic,
  Phys.\ Rev.\ D {\bf 80}, 124027 (2009)
  doi:10.1103/PhysRevD.80.124027
  [arXiv:0906.3250 [hep-th]].

\bibitem{Lopez:1988gt} 
  C.~A.~Lopez,
  Phys.\ Rev.\ D {\bf 38}, 3662 (1988).
  doi:10.1103/PhysRevD.38.3662

\bibitem{Cohen:1968}
  J.~M.~Cohen,
  Phys.\ Rev.\ {\bf173}, 1258 (1968).

\bibitem{Greenwood:2015xwa} 
  E.~Greenwood,
  Phys.\ Lett.\ B {\bf 756}, 365 (2016)
  doi:10.1016/j.physletb.2016.03.041
  [arXiv:1510.05055 [gr-qc]].
  
\bibitem{Maldacena:1997zz} 
  J.~M.~Maldacena,
  AIP Conf.\ Proc.\  {\bf 484}, 51 (1999).
  doi:10.1063/1.59653

\bibitem{Hartnoll:2008vx} 
  S.~A.~Hartnoll, C.~P.~Herzog and G.~T.~Horowitz,
  Phys.\ Rev.\ Lett.\  {\bf 101}, 031601 (2008)
  doi:10.1103/PhysRevLett.101.031601
  [arXiv:0803.3295 [hep-th]].

\bibitem{Ipser:1983db} 
  J.~Ipser and P.~Sikivie,
  Phys.\ Rev.\ D {\bf 30}, 712 (1984).
  
\bibitem{Hendi:2010px} 
  S.~H.~Hendi,
  Eur.\ Phys.\ J.\ C {\bf 71}, 1551 (2011)
  doi:10.1140/epjc/s10052-011-1551-3
  [arXiv:1007.2704 [gr-qc]].

\bibitem{Hendi:2012zz} 
  S.~H.~Hendi,
  JHEP {\bf 1203}, 065 (2012)
  doi:10.1007/JHEP03(2012)065
  [arXiv:1405.4941 [hep-th]].





\end{thebibliography}
\end{document}